\documentclass[a4paper,11pt]{article}
\pdfoutput=1 % if your are submitting a pdflatex (i.e. if you have
\usepackage{jcappub}
\usepackage[utf8]{inputenc}  % for details on the use of the package, please
\usepackage[T1]{fontenc} % if needed
\usepackage{float}% place figure H
\usepackage{xcolor}

\usepackage{caption}
\usepackage{subcaption}

\title{\boldmath Structure Formation in Non-local Bouncing Models}

\author[a]{D. Jackson,\note{Corresponding author.}}
\author[b,1]{R. Bufalo,}
\affiliation[a]{Instituto de Física Teórica, IFT-UNESP,\\R. Dr. Bento Teobaldo Ferraz, 271, Várzea da Barra Funda, São Paulo - SP, 01140-070, Brazil}
\affiliation[b]{Departamento de F\'isica, Universidade Federal de Lavras\\Caixa Postal 3037, 37200-900 Lavras, MG, Brazil}
\emailAdd{dimas.jackson@unesp.br}
\emailAdd{rodrigo.bufalo@ufla.br}
\abstract{
In this study, we investigate the growth of structures within the Deser-Woodard nonlocal theory and extend it to various bouncing cosmology scenarios. Our findings show that the observable structure growth rate, $f\sigma_8$, in a vacuum-dominated universe is finite within the redshift range of $0<z<2$, contrary to previous literature. Although $f\sigma_8$ exhibits no divergences, we observe a slight difference between the evolution of the $\Lambda$CDM and the non-local DW II models. Regarding structure formation in bouncing cosmologies, we evaluate the evolution of $f\sigma_8$ near the bouncing point. Among the different bouncing cases we explore, the oscillatory bounce and pre-inflationary asymmetrical bounce demonstrate a physical profile where the growth rate begins as a small perturbation in the early epoch and increases with inflation, which can be regarded as the seeds of large-scale structures. These findings are significant because they shed light on the growth of seed fluctuations into cosmic structures resulting from non-local effects.}

\begin{document}
\maketitle
\flushbottom

\section{Introduction}
\label{sec:intro}

Despite the fact that more than two decades have passed since the seminal discovery of the accelerated expansion of the Universe \cite{SupernovaSearchTeam:1998fmf,SupernovaCosmologyProject:1998vns}, and that it dominates the Universe’s energy budget and pushes galaxies away at an accelerated pace, the physics mechanism behind it is unclear and still under debate.
The minimal modification of the Einstein gravity in order to handle the accelerated
expansion of current universe is known as the standard cosmological model, or
$\Lambda$CDM.
This model does not change the geometric terms
of Einstein field equation, rather it introduces an extra and assumptive component of matter, called dark energy,  in the form of a cosmological constant $\Lambda$, which is ultimately interpreted as the energy density of the vacuum.

Although the $\Lambda$CDM model possesses a simple structure, and is formally and observationally consistent model, it carries some unsolved puzzles.
In the context of the accelerated expansion of the Universe, we have the coincidence problem: $\Lambda$CDM can not explain why the accelerated phase in the expansion began only recently in the cosmological time. 
Consequently, in order to describe some of these unsolved puzzles a wealth of alternative, more complicated cosmological models are continuously developed and proposed by either changing the matter content (dark energy models) or modified gravity (modify Einstein-Hilbert action to provide extra geometric terms in field equation).

Several modified gravitational theories were proposed as attempts to generalize  Einstein's gravitational theory, usually involving addition of new degrees of freedom.
This can be reached by the insertion of new fields, considering a different geometrical framework or even by enforcing a symmetry principle.
Typically, these new models are required to emulate the background expansion history of the universe given by $\Lambda$CDM, well supported by the data.
The imposition of this condition is called the reconstruction problem \cite{Saini:1999ba,Esposito-Farese:2000pbo}.
Once this step is fulfilled, then one can observationally distinguish among models by looking at their predictions beyond the background, such as solar system tests and the structure formation in the universe \cite{hawking1973large,liddle2000,white20210}. 
It is precisely within the implications of modified gravity models that our interest lies, in special examining whether bouncing cosmologies \cite{Bojowald:2001xe,novello2008bouncing,Ashtekar:2011ni,battefeld2015critical,Brandenberger:2017ni,Nojiri:2017ni} produce physically well-behaved patterns within the context of formation of large scale structures.

An approach to modify the GR inspired by infrared (IR) quantum corrections  are the \textit{non-local theories}, initially proposed in \cite{dalvit1994running, wetterich1998effective}.
In the ref.~\cite{dalvit1994running} the effective equations for the gravitational fields were obtained using a non-local approximation for the quantum effective action, and it was obtained quantum corrections to the newtonian potential.
In contrast, in ref.~\cite{wetterich1998effective} was proposed the addition of a term proportional to $R\square^{-1}R$ to the Einstein-Hilbert action, using a pure phenomenological approach.
This kind of non-local terms involving inverse powers of the d’Alembertian appear in the IR limit
of the quantum effective action \cite{Barvinsky:1985rb,Buchbinder:1992rb,Mukhanov:2007rb,Shapiro:2008sf}.
The issues of causality, domain of validity and boundary conditions in non-local classical and quantum field theories have been discussed \cite{belgacem2018nonlocal,Capozziello:2022lic}, and in both cases, physically viable models can be constructed.
In recent years, we have seen a great interest in phenomenological aspects of nonlocal gravity models \cite{deser2007nonlocal,barvinsky2012serendipitous,maggiore2014nonlocal,belgacem2018nonlocal,belgacem2019testing,Capozziello:2021bki,Capozziello:2022lic,Bouche:2022jts,Capozziello:2022zoz,Bajardi:2022ypn,jackson2022nonlocal}.

Our point of interest is the Deser-Woodard improved model \cite{deser2019nonlocal}, inspired by quantum effective action corrections, which was further elaborated than their previous model \cite{deser2007nonlocal} in order to fully satisfy the screening mechanism and also to reproduce the late time accelerated expansion of $\Lambda$CDM via the reconstruction procedure, without the necessity of a cosmological constant.
\footnote{The original DW I model, based on the Lagrangian   $\mathcal{L}=R f \left(\square^{-1} R \right) $ where $f$ is an algebraic function, has been shown to be inconsistent, since it failed a decisive test by not satisfying the screening mechanism to avoid non-local effects in Solar System scale, violating thus observational constraints \cite{belgacem2018nonlocal}.}
This improved model, which we will call DW II, minimally modifies the Einstein-Hilbert action, by the presence of a algebraic function of a non-local operator
\begin{equation}
\mathcal{L}=\frac{1}{16\pi G}R\left[1+f(Y)\right],\label{eq:LDW}
\end{equation}
where
\begin{subequations}
	\begin{align}
	Y & =\square^{-1}g^{\mu\nu}\partial_{\mu}X\partial_{\nu}X\,, \\
	X & =\square^{-1}R\,.\label{eq:X}
	\end{align}
\end{subequations}
Here $\square=g^{\mu\nu}D_{\mu}D_{\nu}$ is the covariant d'Alembertian operator and $R$ is the curvature scalar.
Actually, the term $g^{\mu\nu}\partial_{\mu}X\partial_{\nu}X$ is negative in the Solar System scale and positive in the cosmological scale, which makes viable a screening effect.

The two Deser-Woodard models have been already examined in the context of structure formation 	\cite{Park:2012cp,Dodelson:2013sma,Nersisyan:2017mgj,ding2019structure}.
The authors studied the growth rate $f\sigma_8$ predicted by the DW model in $\Lambda$CDM background, and found that the models lead to a good agreement with the Redshift-space distortions observations (RSD), which is known to provide a big database for testing modified gravity models.
However, some discrepancies have been found among these analyzes.

We therefore revisit the analysis of structural growth rate of the Universe for the DW II model in order to shed some light in its issues, and show how our results disagree with those in \cite{ding2019structure}.
Our analysis shows a growth rate $f\sigma_8$ continuous for $0<z<2$, see Fig.~\ref{fig:fsigma8}, while $f\sigma_8$ has a prominent discontinuity in \cite{ding2019structure}.
Furthermore, we also extend the bouncing solutions examined at the level of background cosmology in the DW II model \cite{jackson2022nonlocal} to the early time perturbations, by discussing bouncing models in the context of structure formation.
The most interesting result is that in some bouncing universes we conclude that the growth of seed fluctuations into cosmic (large scale) structure can be ascribed to non-local effects.

The main interest of the present work is to analyze the perturbative growth of structures in the $\Lambda$CDM for the DW II model and extend it to five different bouncing cosmology models: symmetric bounce \cite{Cai:2012va,Cai:2013vm,bamba2014bounce},
oscillatory bounce \cite{novello2008bouncing,Cai:2012va,Steinhardt:2001st},
 matter bounce \cite{Singh:2006im,Wilson-Ewing:2012lmx}, 
 finite time singularity model \cite{Barrow:2015ora,Nojiri:2015fra,Odintsov:2015zza,Oikonomou:2015qha}
 and   pre-inflationary asymmetric bounce \cite{odintsov2022preinflationary}.
The paper is organized as follows: in Sec.~\ref{sec:imp} we review the main features of the DW II model, in special, in the reconstruction process to determine the distortion function $f(Y)$ that emulates the $\Lambda$CDM cosmology.
In Section \ref{sec:perturb} we workout the cosmological (time) perturbation, considering scalar perturbations over a flat FLRW geometry background, and evaluate numerically the solution to the contrast density of matter and its physical observable, the structural growth rate $f\sigma_8$.
We discuss these results in order to highlight some possible causes to the disagreement with those in \cite{ding2019structure}.
In Section \ref{sec:grow} we examine the evolution of the growth rate $f\sigma_8$ for the aforementioned bounce models.
For the cases of oscillatory and asymmetrical bouncing universes, we find that they render physically acceptable patterns for the $f\sigma_8$, which allow us to conclude that the growth of seed fluctuations into cosmological structures can be ascribed to non-local effects.
At last, we present our final remarks and perspectives in Sec.~\ref{sec:conc}.

\section{Reconstruction procedure and background equations}
\label{sec:imp}

In this section we shall describe the main aspects of the analysis regarding the zeroth order perturbative field equations (based on the action \eqref{eq:LDW}), which is based in the reconstruction process in order to obtain the solution to the distortion function $f(Y)$.
The first step in the reconstruction process is to localize the action, which can be achieved by the introduction of two auxiliary scalar fields $U$ and $V$ as Lagrange multipliers in equation \eqref{eq:LDW}, resulting into
\begin{equation}\label{eq:localDW}
\mathcal{L}=\frac{1}{16\pi G}\left[R\left(1+f(Y)+U\right)+g^{\mu\nu}E_{\mu\nu}\right],
\end{equation}
in which we have introduced $E_{\mu\nu} = \partial_{\mu}U\partial_{\nu}X+\partial_{\mu}V\partial_{\nu}Y+V\partial_{\mu}X\partial_{\nu}X$, by means of notation. Hence, by considering $X,Y,U$ and $V$ as four independent scalar fields, the action $S=\int d^4x\mathcal{L}$ is regarded as local.

One can observe from \eqref{eq:localDW}, obtained after localization procedures, that non-local terms result as
effective scalar fields. 
From a phenomenological point of view, this means that non-local corrections give rise to effective lengths and masses which could alleviate several shortcomings of General Relativity at UV and IR scales, intrinsically related with regularization and renormalization of gravitational effective action.

Another important aspect of the model is that the fields $X,Y,U$ and $V$ in equation \eqref{eq:localDW} are subject to retarded boundary conditions \cite{deser2019nonlocal}, which require that all the fields and their first time derivatives vanish in an initial value surface.
In summary, unless these boundary conditions are satisfied by all the auxiliary scalars fields, unwanted new degrees of freedom would arise, these are known as ghosts (since they have negative kinetic terms) \cite{deser2019nonlocal}.
Hence, retarded boundary conditions will be used throughout our analysis.

Varying the action in respect to each of these fields, result into the following set of (constraint) equations
\begin{subequations}
\label{eq:constraints}
\begin{align}
R-\square X &=0\,,\label{eq:XDWII}
\\
g^{\mu\nu}\partial_{\mu}X\partial_{\nu}X-\square Y &=0\,,\label{eq:Y}
\\
2D^{\mu}\left(VD_{\mu}X\right) + \square U &=0\,,\label{eq:U}
\\
R\frac{\partial f(Y)}{\partial Y} -\square V &=0\,.\label{eq:V}
\end{align}
\end{subequations}
The gravitational field equations are obtained by varying the action  \eqref{eq:localDW} in respect to the metric $g^{\mu\nu}$ \footnote{The indices in parenthesis denote the symmetric part $
E_{(\mu\nu)}=\frac{1}{2}\left(E_{\mu\nu}+E_{\nu\mu}\right)\,.$}.
\begin{equation}
\left(G_{\mu\nu}-D_{\mu}D_{\nu}+g_{\mu\nu}\square\right)\left(1+U+f(Y)\right)+E_{(\mu\nu)}-\frac{1}{2}g_{\mu\nu}g^{\rho\sigma}E_{\rho\sigma}=8\pi GT_{\mu\nu}\,,\label{eq:DW2field}
\end{equation}
where the energy momentum tensor $T_{\mu\nu}=\left(\rho+p\right)u_{\mu}u_{\nu}+pg_{\mu\nu}$ corresponds to the usual baryonic matter and does not include the dark energy source term 
 
This non-local model is known to reproduce the current accelerated expansion  of the universe without cosmological constant when the \textit{non-local distortion function} $f(Y)$ satisfies
 \begin{equation}
 	f(Y) \sim e^{1.1(Y+16.7)} \,.
 	\label{eq:fit}
 \end{equation}
 This expression is an exponential fit to the numerical solution obtained trough the reconstruction process \cite{deser2019nonlocal}, which consists in requiring that the Friedmann equations of General Relativity should be satisfied by the DW II model. 

Since we wish to analyze some bouncing universes within the DW II model at perturbative level, which corresponds in examine the behavior of the distortion function $f(Y)$ trough the reconstruction process under the influence of bouncing universes, we shall revise aspects of the reconstruction process regarding the (zeroth order perturbative) field equations necessary to obtain \eqref{eq:fit}.

 We start the reconstruction procedure by expanding the field equations \eqref{eq:DW2field} over the Friedmann-Lemaître-Robertson-Walker (FLRW) background:
 \begin{equation}
	ds^2=dt^2-a^2(t)dx_i dx^i\,.
 \label{eq:metric}
 \end{equation}
 This metric can also be seen as the zeroth order perturbative metric in the newtonian gauge.  
The d'Alembertian operator acting on a scalar function $W(t)$, which depends only on time, is written as
 \begin{align}
 \square W(t) &= d_{t}^{2}W(t)+3H d_{t}W(t)\,,\label{eq:BOXW}
 \end{align}
 where $H=\frac{\dot{a}}{a}$ is the Hubble parameter.
 Therefore, the zeroth order field equations (00) and (ij) components \eqref{eq:DW2field} are, respectively, given by
 \begin{subequations}
 \begin{align}
 \left(3H^{2}+3Hd_t\right)\left(1+U+f(Y)\right)+\frac{1}{2}\left(\dot{X}\dot{U}+\dot{Y}\dot{V}+V\dot{X}^{2}\right) & =8\pi G\rho\,,\label{eq:00}
 \\
 -\left[2\dot{H}+3H^{2}+d_t^{2}+2Hd_t\right]\left(1+U+f(Y)\right)+\frac{1}{2}\left(\dot{X}\dot{U}+\dot{Y}\dot{V}+V\dot{X}^{2}\right) & =8\pi Gp\,.\label{eq:11}
 \end{align}
 \end{subequations}
 Furthermore, subtracting the equations \eqref{eq:00} and \eqref{eq:11} we find a differential equation for the function $F(t)\equiv1+U(t)+f[Y(t)]$, which is cast as
 \begin{equation}
 \left[2\dot{H}+6H^{2}+d_t^{2}+5Hd_t\right]F(t)=8\pi G\left(\rho-p\right)\,.\label{eq:FNL}
 \end{equation}
 For the reconstruction process analysis, it is convenient to parametrize the (time dependence of the) field equations in terms of the $e$-folding time, $N=\ln a_0/a$ (with $a_0 =1$), so that $f(Y)$ can be solved independently of a particular form of the scale factor \footnote{In this case is required only that the universe remains expanding, increasing in size by a factor of $e$, $N$ times. In Section \ref{sec:grow}, on the other hand, we will discuss some bouncing universes, i.e. the collapse and re-expansion.}.
Since we want to reconstruct the accelerated expansion, the Friedmann equations from $\Lambda$CDM are used as source
\begin{subequations}
	\begin{align}
H^2& =H_0^2 \left(\Omega_{M} e^{3N}+\Omega_{R}e^{4N}+\Omega_\Lambda \right)\,,
\label{eq:Hquad}
	\end{align}
	\begin{align}
8\pi G \rho &=3H_0^2\left( \Omega_{R}e^{4N}+\Omega_{M}e^{3N} \right) \,,
\label{eq:rho}
	\end{align}
	\begin{align}
8\pi G p &=3H_0^2 \dfrac{\Omega_{R}}{3}e^{4N} \,.
\label{eq:p}
	\end{align}
 \end{subequations}
 The parameters $\Omega_{M}, \Omega_{R}$ and $\Omega_\Lambda $ are respectively the matter, radiation and dark energy fractions of energy density at the present day.
Hence, applying the above changes, equation \eqref{eq:FNL} becomes 
 \begin{align}
 	\left[\partial_{N}^{2}+(\epsilon-5)\partial_{N}+(6-2\epsilon)\right]F(N)=\frac{H_{0}^{2}}{H^{2}}\left(3e^{3N}\ensuremath{\Omega}_{M}+2e^{4N}\Omega_{R}\right)\,,
 	\label{eq:difF}
 \end{align}
  where $\epsilon=\partial_N H / H = -\dot{H}/H^2 $.
%  Actually, our result \eqref{eq:difF} and is slightly different from equation (29) of \cite{deser2019nonlocal}. 
 % {\color{red}Our equation (\ref{eq:difF}) comes from the subtraction of the  equations (\ref{eq:rho}) and (\ref{eq:p}) whereas equation (29) of \cite{deser2019nonlocal} seems not to be the difference between their equations (21) and (22).}
  
 At last, we turn our attention to the auxiliary field equations \eqref{eq:XDWII}-\eqref{eq:V}: we expand them over the background \eqref{eq:metric} and also write them $e$-folding time $N$, yielding to
 \begin{subequations}
 \begin{align}
 12\left(1-\dfrac{\epsilon}{2}\right)\dfrac{H}{H_{0}}e^{-3N}+\partial_{N}\left(\dfrac{H}{H_{0}}e^{-3N}\partial_{N}X\right)	=0\,, \\
 -\left(\partial_{N}X\right)^{2}\dfrac{H}{H_{0}}e^{-3N}+\partial_{N}\left(\dfrac{H}{H_{0}}e^{-3N}\partial_{N}Y\right)	=0\,, \\
 \partial_{N}U+2V\partial_{N}X	=0\,, \\
 2\partial_{N}^{2}V+2\left(\epsilon-3\right)\partial_{N}V+12\left(2-\epsilon\right)\left(2\dfrac{\partial_{N}X}{\partial_{N}Y}V+\dfrac{\partial_{N}F}{\partial_{N}Y}\right)	=0\,,
 \end{align}
 \end{subequations}

\begin{figure}[H]
	\centering
	\includegraphics[scale=0.9]{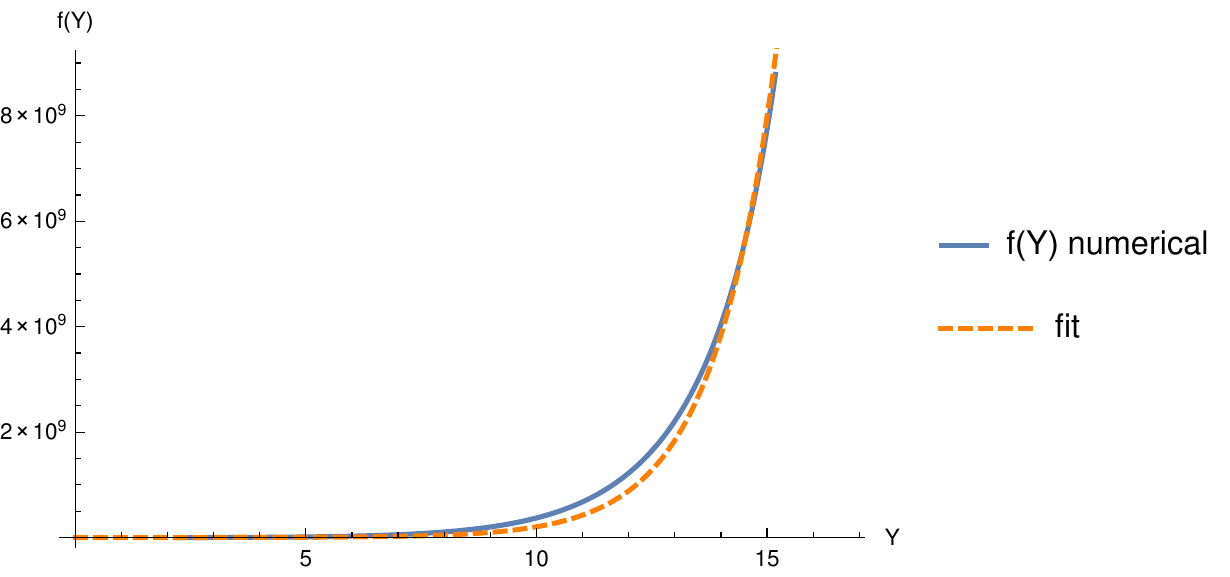}
	\caption{Reconstruction process for the distortion function $f(Y)$ for the $\Lambda$CDM universe.}
	\label{fig:fdeY}
\end{figure}

Therefore, having in hand the solutions for $F$ and $U$, the non-local distortion function can be numerically obtained through the relation $f=F-U-1$.  The exact solution, which emulates the $\Lambda$CDM cosmology, reads
\begin{align}
 	f(Y)\approx e^{0.73 (Y+16.13)}\,.
 	\label{eq:fdeY}
 \end{align}
and the curve fitted to the solution is present in Figure \ref{fig:fdeY}.
On can note that our result \eqref{eq:fdeY} is not precisely the same of the original authors \eqref{eq:fit} (which might which may result from the small difference of the initial conditions), but our distortion function also presents the desired exponential growth at the recent epoch.
 
One last remark is that equations \eqref{eq:00} and \eqref{eq:11} are the zeroth order perturbative field equations (which ultimately led to our eq.~\eqref{eq:fdeY}), and following the analysis developed above we will calculate the first order perturbative equations in the newtonian gauge in the next section.
  
\section{Cosmological Perturbation Theory}
\label{sec:perturb}

In this section we will calculate the perturbative field equations for the DW II model, laying the ground for the application of the reconstruction process for bouncing universes.
We will consider scalar cosmological perturbations, starting from the perturbative metric on the newtonian gauge, which is given by
\begin{equation}
ds^{2}=\left(1+2\Psi(\vec{r},t)\right)dt^{2}-\left(1+2\Phi(\vec{r},t)\right)a^{2}(t)dx^{i}dx_{i}.\label{eq:perturbedFLRW}
\end{equation}
Here $\Psi$ and $\Phi$ are the two gauge invariant perturbation degrees of freedom, also called the Bardeen potentials \cite{liddle2000,white20210,ma1995cosmological}.

The validity of the metric \eqref{eq:perturbedFLRW} in the description of large scale structures have been extensively discussed in the literature: analytical arguments and computational simulations have favoured that this metric provides a good approximation  to the actual metric of the Universe (in the scalar-perturbation sector), encompassing the cosmological solution FLRW and the static Schwarzchild solution, for further details see \cite{ma1995cosmological,belgacem2019testing} and references therein. 

Let us now present some important results and remarks to write down the perturbative field equations, as well some key aspects related with the non-local contributions.
Keeping up to the first-order terms in the perturbation, %the non-null components of the connection are
%\begin{subequations}
%\begin{align}
%\Gamma_{\hphantom{\rho}00}^{0} & =\partial_{t}\Psi\\
%\Gamma_{\phantom{0}ij}^{0} & =a^{2}\left(H+\partial_{t}\Phi+2H\left(\Phi-\Psi\right)\right)\delta_{ij}\\
%\Gamma_{\hphantom{\rho}k0}^{0} & =\partial_{k}\Psi\\
%\Gamma_{\hphantom{\rho}00}^{k} & =a^{-2}\partial^{k}\Psi\\
%\Gamma_{\hphantom{\rho}0j}^{i} & =\left(H+\partial_{t}\Phi\right)\delta_{\phantom{j}j}^{i}\\
%\Gamma_{\phantom{\gamma}jk}^{i} & =-\delta^{il}\partial_{l}\Phi\delta_{jk}+\partial_{j}\Phi\delta_{k}^{\phantom{k}i}+\partial_{k}\Phi\delta_{\phantom{i}j}^{i},
%\end{align}
%\end{subequations}
one founds the $00$ and $(ij)$ components of the Ricci tensor
\begin{subequations}
\begin{align}
R_{00} & =a^{-2}\nabla^{2}\Psi+3H\left(\partial_{t}\Psi-2\partial_{t}\Phi\right)-3\dot{H}-3\partial_{t}^{2}\Phi-3H^{2}\\
R_{ij} & =a^{2}\left[3H^{2}+\dot{H}-H\partial_{t}\left(\Psi-6\Phi\right)+\partial_{t}^{2}\Phi+2\left(3H^{2}+\dot{H}\right)\left(\Phi-\Psi\right)\right]\delta_{ij}\nonumber \\
&\quad-\partial_{i}\partial_{j}\left(\Phi+\Psi\right)-\nabla^{2}\Phi\delta_{ij}.
\end{align}
\end{subequations}
The perturbed Einstein tensor can be cast as
\begin{subequations}
\begin{align}
G_{00}&=3H^{2}+6H\partial_{t}\Phi-2a^{-2}\nabla^{2}\Phi\\
G_{ij}&=a^{2}\left[-\left(3H^{2}+2\dot{H}\right)\left(1+2\Phi-2\Psi\right)-H\partial_{t}\left(2\Psi+6\Phi\right)-2\partial_{t}^{2}\Phi\right]\delta_{ij} \nonumber \\ 
&\quad+\nabla^{2}\left(\Phi+\Psi\right)\delta_{ij}-\partial_{i}\partial_{j}\left(\Phi+\Psi\right),
\label{eq:Gij}
\end{align}
\end{subequations}
which can be identified as $G_{\mu\nu}=\bar{G}_{\mu\nu}+\delta G_{\mu\nu}$, where $\bar{G}_{\mu\nu}$ are the zero-order part and $\delta G_{\mu\nu}$ the first-order perturbation.
On the other hand, the field equations of the DW II model \eqref{eq:DW2field} can be written as $G_{\mu\nu}+\Delta G_{\mu\nu}=8\pi GT_{\mu\nu}$, in which the symbol $\Delta$ denotes the non-local correction and must not be confused with the perturbative correction, represented by $\delta$.
Therefore, we found the non-local contribution of the DW II model
\begin{align}
\Delta G_{\mu\nu}=\left(U+f\left(Y\right)\right)G_{\mu\nu}+\left(g_{\mu\nu}\square-D_{\mu}D_{\nu}\right)\left(U+f\left(Y\right)\right)+E_{(\mu\nu)}-\frac{1}{2}g_{\mu\nu}g^{\rho\sigma}E_{\rho\sigma}.
\label{eq:DeltaGNL}
\end{align}
In order to obtain the perturbative field equations, we decompose the perturbed auxiliary fields $(X,Y,U,V)$, into the background term and the perturbation,
\begin{subequations}
	\begin{align}
U(\vec{r},t)=U_{c}\left(t\right)+\delta U\left(\vec{r},t\right)\,,\quad X(\vec{r},t)=X_{c}\left(t\right)+\delta X\left(\vec{r},t\right)\,,
\\
V(\vec{r},t)=V_{c}\left(t\right)+\delta V\left(\vec{r},t\right)\,,\quad Y(\vec{r},t)=Y_{c}\left(t\right)+\delta Y\left(\vec{r},t\right)\,,
\label{eq:pertufields}
\end{align}
\end{subequations}
where the subscript $c$ denotes that the fields are evaluated in the time dependent cosmological background.
The spatial dependence of the (perturbed) fields $X,Y,U$ and $V$ can be readily understood from the fact that the perturbative potentials introduced in the metric \eqref{eq:perturbedFLRW} depends on $\vec{r}$. 
Moreover, in our analysis of the perturbed field equations, we will also need the perturbative expression of the d'Alembertian operator, which in first-order reads
\begin{align}
\square X(\vec{r},t) & =\left(1+2\Psi\right)^{-1}\partial_{t}^{2}X-a^{-2}\left(1+2\Phi\right)^{-1}\nabla^{2}X\cr
&\quad -\partial_{t}\Psi\partial_{t}X+\left(3H+3\partial_{t}\Phi-6H\Psi\right)\partial_{t}X \cr
& \quad+a^{-2}\left[\partial_{x}\left(\Psi+\Phi\right)\partial_{x}X+\partial_{y}\left(\Psi+\Phi\right)\partial_{y}X+\partial_{z}\left(\Psi+\Phi\right)\partial_{z}X\right].\label{eq:boxXpertu}
\end{align}
Finally, by replacing the results \eqref{eq:boxXpertu} and \eqref{eq:pertufields} in \eqref{eq:DeltaGNL}, and after some algebraic manipulations,  one can find the perturbed non-local correction of the 00 Einstein equation
\begin{align}
\delta\Delta G_{00} & =\left[3\partial_{t}\Phi\partial_{t}+6H\partial_{t}\Phi-\partial_{t}^{2}-2\dfrac{\nabla^{2}}{a^{2}}\Phi\right]\left(U_{c}+f\right) \cr
& \quad+\left[-\dfrac{\nabla^{2}}{a^{2}}+3H\partial_{t}+3H^{2}\right]\left(\delta U+\dfrac{df}{dY}\delta Y\right)\nonumber \\
& \quad+\dfrac{1}{2}\left(\dot{X_{c}}\dot{\delta U}+\dot{U_{c}}\dot{\delta X}+\dot{Y_{c}}\dot{\delta V}+\dot{V_{c}}\dot{\delta Y}+2V_{c}\dot{X}_{c}\dot{\delta X}+\delta VX_{c}^{2}\right).
\end{align}
To complete the perturbative field equations we also expand the stress-energy tensor
\begin{align}
\delta T_{00}  =\rho_{c}\dfrac{\delta\rho}{\rho_{c}} \equiv \rho_{c}\delta
\end{align}
where the matter density parameter (also called density contrast) is defined by $\delta\equiv\dfrac{\delta\rho}{\rho_{c}}$. 

Our perturbative analysis of the reconstruction process (for bouncing universes) takes place in the Fourier space: we consider (spatial) plane wave solutions for the perturbative modes; this implies that the spatial Laplacian operator is rewritten as $-\nabla^{2}\to k^{2}$.
In this approach, we will restrict ourselves to the sub-horizon limit ($k\gg\dot{a}$ or $\frac{k}{aH}\gg1$), i.e., in which the spatial derivatives are more relevant than the time derivatives.
Physically speaking, this means that we are only considering perturbative modes with wavelength $k^{-1}$ much less than the Hubble distance $(aH)^{-1}$.
Hence, in the sub-horizon limit \footnote{All expressions henceforth are computed in the sub-horizon limit.}, the first-order part of the 00 field equation assumes a reduced form
\begin{equation}
2\Phi\left(1+U_{c}+f\right)+\delta U+\dfrac{df}{dY}\delta Y=8\pi G\dfrac{a^{2}}{k^{2}}\rho\delta.\label{eq:pert_field_00}
\end{equation}
For the $(ij)$ components, we obtain from \eqref{eq:Gij} the perturbed Einstein tensor 
\begin{equation}
\delta G_{ij}=\left(-a^{2}k^{2}\delta_{ij}+k_{i}k_{j}\right)\left(\Phi+\Psi\right).
\label{eq:deltaGij}
\end{equation}
Furthermore, the expression \eqref{eq:deltaGij} can be rewritten in a more convenient form by acting with the projection operator $\left(\dfrac{k^{i}k^{j}}{k^{2}}-\dfrac{1}{3}\delta^{ij}\right)$ \cite{ma1995cosmological}, which yields
\begin{align} \label{eq_pert_local}
\left(\dfrac{k^{i}k^{j}}{k^{2}}-\dfrac{1}{3}\delta^{ij}\right)\delta G_{ij} =\dfrac{2}{3}k^{2}\left(\Phi+\Psi\right),
\end{align}
%This projection operator is known to relate the longitudinal part of the perturbative metric in the sincronous gauge to the newtonian gauge \cite{ma1995cosmological}.
Therefore, the perturbative expansion of the non-local part of the $(ij)$ components is written as
\begin{equation}\label{eq_pertij}
\left(\dfrac{k^{i}k^{j}}{k^{2}}-\dfrac{1}{3}\delta^{ij}\right)\delta\Delta G_{ij}=\dfrac{2}{3}k^{2}\left[\left(U_{c}+f\right)\left(\Phi+\Psi\right)+\left(\delta U+\dfrac{df}{dY}\delta Y\right)\right]. 
\end{equation}

With the result \eqref{eq_pertij} we have concluded the perturbative analysis of the metric part of the $(ij)$ field equations, we now turn our attention to the source term.
In our metric signature, the variation of the stress-energy tensor is
\begin{equation}
T_{\phantom{i}j}^{i}=-p\delta_{\phantom{i}j}^{i}+\Sigma_{\phantom{i}j}^{i},
\end{equation}
where $\Sigma_{\phantom{i}j}^{i}\equiv T_{\phantom{i}j}^{i}-\dfrac{\delta_{\phantom{i}j}^{i}}{3}T_{\phantom{i}k}^{k}$ is the traceless part of the tensor $T_{ij}$.
It is worth mention that in the case where the source	 consists of
radiation and non-relativistic matter, we have a vanishing anisotropic stress tensor $\Sigma_{\phantom{i}j}^{i} \simeq 0$.
Moreover, defining a anisotropic stress $\sigma$ such that,
\begin{equation}
\left(\rho+p\right)\sigma\equiv\left(\dfrac{k^{i}k^{j}}{k^{2}}-\dfrac{1}{3}\delta^{ij}\right)\Sigma_{ij},
\end{equation}
we find
\begin{align} \label{eq_pert_stress}
\left(\dfrac{k^{i}k^{j}}{k^{2}}-\dfrac{1}{3}\delta^{ij}\right)\delta T_{ij} =\left(\rho+p\right)\sigma.
\end{align}

With these results eqs.~\eqref{eq_pert_local}, \eqref{eq_pertij} and  \eqref{eq_pert_stress},  one can calculate the longitudinal component of the $(ij)$ field equations, which at the leading order in the limit $k\gg aH$, is given by
\begin{align}
%\left(\dfrac{k^{i}k^{j}}{k^{2}}-\dfrac{1}{3}\delta^{ij}\right)\left(\delta G_{ij}+\delta\Delta G_{ij}\right) & =8\pi G\left(\dfrac{k^{i}k^{j}}{k^{2}}-\dfrac{1}{3}\delta^{ij}\right)\delta T_{ij}\\
\dfrac{2}{3}k^{2}\left(\Phi+\Psi\right)+\dfrac{2}{3}k^{2}\left[\left(U_{c}+f\right)\left(\Phi+\Psi\right)-\left(\delta U+\dfrac{df}{dY}\delta Y\right)\right] & =\left(\rho+p\right)\sigma.
\end{align}
Finally, in the (late time) epoch when the relativistic contribution is small, we can neglect the contribution coming from the anisotropic stress $\sigma\approx0$.
Thus, we get 	
\begin{equation}
\left(\Phi+\Psi\right)\left(1+U_{c}+f\right)-\left(\delta U+\dfrac{df}{dY}\delta Y\right)=0.\label{eq:pert_field_ij}
\end{equation}
The equations \eqref{eq:pert_field_00} and \eqref{eq:pert_field_ij}
comprise the perturbative (metric) field equations of the DW II non-local gravity, in the sub-horizon limit.
However, in order to fully determine the potentials introduced in the metric \eqref{eq:perturbedFLRW} and complete our reconstruction process, it is necessary to obtain the perturbative expansion of the auxiliary scalar fields introduced in the action \eqref{eq:localDW}.

%%%%%%%%%%%%%%%%%%%%%%%%%%%%%%%%%%%%%%%%%%%%%%%%%%%%%%%%%
%%%%%%%%%%%%%%%%%%%%%%%%%%%%%%%%%%%%%%%%%%%%%%%%%%%%%%%%%

\subsection{Perturbative expansion of the auxiliary fields}
\label{subsec:perturbativas}

We shall now solve the perturbative equations for the auxiliary fields eq.~\eqref{eq:constraints}, which
together with the metric field equations \eqref{eq:pert_field_00} and \eqref{eq:pert_field_ij}, form the set of six equations for the six undetermined variables $\left( \delta X,\delta Y,\delta U,\delta V,\Phi,\Psi\right)$.
Hence, the full set of first-order perturbed equations is explicitly written as:
\begin{subequations}
\begin{align}
8\pi G\dfrac{a^{2}}{k^{2}}\rho\delta & =2\Phi\left(1+U_{c}+f\right)+\delta U+\dfrac{df}{dY}\delta Y,\\
\delta U+\dfrac{df}{dY}\delta Y & =\left(\Phi+\Psi\right)\left(1+U_{c}+f\right),\\
\delta X & =-2\left(2\Phi+\Psi\right),\\
\delta Y & =0,\\
\delta U & =2V_{c}\delta X,\\
\delta V & =-2\left(2\Phi+\Psi\right)\frac{\partial f}{\partial Y}.
\end{align}
\end{subequations}
Eliminating $\delta Y$ e $\delta U$ in the two first equations, results into
\begin{subequations}
\begin{align}
8\pi G\dfrac{a^{2}}{k^{2}}\rho\delta & =2\Phi\left(1+U_{c}+f\right)-4V_{c}\left(2\Phi+\Psi\right)\\
-4V_{c}\left(2\Phi+\Psi\right) & =\left(\Phi+\Psi\right)\left[1+a^{2}\left(U_{c}+f\right)\right].
\end{align}
\end{subequations}
Solving algebraically for the potentials $\Phi$ and $\Psi$, it yields
\begin{subequations}
\begin{align}
\Phi & =\dfrac{4\pi Ga^{2}\rho\delta}{k^{2}\left(1+f_c+U_{c}\right)}\dfrac{\left(1+f_c+U_{c}+4V_{c}\right)}{\left(1+f_c+U_{c}+2V_{c}\right)},\label{eq:phi-1}\\
\Psi & =-\dfrac{4\pi Ga^{2}\rho\delta}{k^{2}\left(1+f_c+U_{c}\right)}\dfrac{\left(1+f_c+U_{c}+8V_{c}\right)}{\left(1+f_c+U_{c}+2V_{c}\right)},\label{eq:psi}
\end{align}
\end{subequations}
where $f_c\equiv f(Y_{c}).$
Hence, we observe that the potentials $\Phi$ and $\Psi$ are fully determined in terms of the auxiliary fields evaluated in the cosmological background. 

For the analysis of the matter density perturbation (discussed below), it is convenient  to separate the matter and the radiation contributions to the energy density, i.e. $\rho=\rho_{R}+\rho_{M}$, where
\begin{subequations}
\begin{align}
\rho_{R} & =\rho_{0R}a^{-4}\\
\rho_{M} & =\rho_{0M}a^{-3}.
\end{align}
\end{subequations}
Thus, we can rewrite equations \eqref{eq:phi-1} and \eqref{eq:psi} in terms of the density parameter $\Omega^{0}\equiv\dfrac{8\pi G}{3H_{0}^{2}}\rho_{0}$,
\begin{subequations}
\begin{align}
\Phi & =\dfrac{3H_{0}^{2}\left(\Omega_{R}^{0}a^{-2}\rho_{0R}\delta_{R}+\Omega_{M}^{0}a^{-1}\rho_{0M}\delta_{M}\right)}{2k^{2}\left(1+f_c+U_{c}\right)}\dfrac{\left(1+f_c+U_{c}+4V_{c}\right)}{\left(1+f_c+U_{c}+2V_{c}\right)},\\
\Psi & =-\dfrac{3H_{0}^{2}\left(\Omega_{R}^{0}a^{-2}\rho_{0R}\delta_{R}+\Omega_{M}^{0}a^{-1}\rho_{0M}\delta_{M}\right)}{2k^{2}\left(1+f_c+U_{c}\right)}\dfrac{\left(1+f_c+U_{c}-8V_{c}\right)}{\left(1+f_c+U_{c}+2V_{c}\right)}.
\end{align}
\end{subequations}
As discussed above,  in the sub-horizon limit $k\gg aH$, the non-relativistic matter is more relevant  than the radiation one. Therefore, the potentials are solely expressed in terms of the matter density perturbation
\begin{subequations}
\begin{align}
\Phi & =\dfrac{3H_{0}^{2}}{2ak^{2}}\dfrac{\left(1+f_c+U_{c}+4V_{c}\right)}{\left(1+f_c+U_{c}+2V_{c}\right)}\dfrac{\Omega_{M}^{0}\rho_{0M}\delta_{M}}{\left(1+f_c+U_{c}\right)},\label{eq:psi-1}\\
\Psi & =-\dfrac{3H_{0}^{2}}{2ak^{2}}\dfrac{\left(1+f_c+U_{c}+8V_{c}\right)}{\left(1+f_c+U_{c}+2V_{c}\right)}\dfrac{\Omega_{M}^{0}\rho_{0M}\delta_{M}}{\left(1+f_c+U_{c}\right)}.\label{eq:psi-2}
\end{align}
\end{subequations}
Some remarks about our results for the potentials \eqref{eq:psi-1} and \eqref{eq:psi-2}:
once the matter density contrast $\delta_{M}$ is obtained, the potentials $\Phi$ and $\Psi$ are determined.
In addition, we shall use our result for $\delta_{M}$ to compare it with $f\sigma_8$ data.
At last, we will discuss the effects of bouncing universes in the profile of the matter density contrast $\delta_{M}$, and consequently in the potentials.
These aspects will be analyzed in the next sections.

One last remark is that our expressions are similar to those found in \cite{ding2019structure}, except for a small difference: the term $2V_{c}$ in the denominator appears as $6V_{c}$ in equations (2.35) and (2.37) of \cite{ding2019structure}. 

%%%%%%%%%%%%%%%%%%%%%%%%%%%%%%%%%%%%%%%%%%%%%%%%%%%%%%%%%
%%%%%%%%%%%%%%%%%%%%%%%%%%%%%%%%%%%%%%%%%%%%%%%%%%%%%%%%%

\subsection{Structural Growth in Non-local Expanding Universe}

In order to examine the implications of bouncing universes in non-local gravity models over the structure formation, we must first determine the solution for the density $\delta_{M}$.
With this motivation, we establish here the differential equation for the matter density contrast $\delta_{M}$ and obtain its numerical solution.

The differential equation for the matter density contrast $\delta_{M}$ can be obtained by using the conservation law for the stress-energy tensor, $D_{\mu}T_{\phantom{0}\nu}^{\mu}=0$.
Consider the perturbation of the perfect fluid stress-energy tensor \cite{weinberg1972gravitation}, 
\begin{subequations}
\begin{align}
T_{\phantom{0}0}^{0} & =\rho_{c}+\delta\rho, \\
T_{\phantom{0}i}^{0} & =\left(\rho_{c}+p_{c}\right)v_{i}=-a^2T_{\phantom{0}0}^{i}, \\
T_{\phantom{0}i}^{j} & =-\left(p+\delta p\right)\delta_{\phantom{0}i}^{j}+\Sigma_{\phantom{0}i}^{j},
\end{align}
\end{subequations}
where $v_i \equiv dx_i/d\tau$ is the coordinate velocity.
The $\nu=0$ component of the conservation law, at first-order approximation, provides
\begin{equation}
\dot{\delta\rho}-a^{-2}\partial_{j}\left(\rho_{c}+p_{c}\right)v^{j}+3H\left(\delta\rho+\delta p\right)+3\dot{\Phi}\left(\rho_{c}+p_{c}\right)=0.
\end{equation}
Moreover, using the equation of state $p_{c}=w\rho_{c}$, as well as the fluid sound speed $\dfrac{\delta p}{\delta\rho}=c_{s}^{2}$, we obtain
\begin{equation}
\dot{\delta}+\dfrac{\dot{\rho_{c}}}{\rho_{c}}\delta+\left(3\dot{\Phi}-a^{-2}\vec{\nabla}\cdot\vec{v}\right)\left(1+w\right)+3H\left(1+c_{s}^{2}\right)\delta=0.\label{eq:ordem1}
\end{equation}
We can also use the zeroth order equation, $\dot{\rho_{c}}=-3H\left(\rho_{c}+p_{c}\right)=-3H\left(1+w\right)\rho_{c}$,
to simplify the relation \eqref{eq:ordem1} as
\begin{equation}
\dot{\delta}+\left(3\dot{\Phi}-a^{-2}\vec{\nabla}\cdot\vec{v}\right)\left(1+w\right)+3H\left(c_{s}^{2}-w\right)\delta=0.\label{eq:nu0}
\end{equation}
On the other hand, for the spatial components $\nu=i$ it reads
\begin{equation}
\dot{v}_{i}=3H\dfrac{\dot{p}_{c}}{\dot{\rho}_{c}}v_{i}+\dfrac{\partial_{i}\delta p}{\left(p_{c}+\rho_{c}\right)}+\partial_{i}\Psi. \label{eq:nu1}
\end{equation}
%then, summing up the three vector components, follow that
%\begin{equation}
%\dot{\vec{v}}=3H\dfrac{\dot{p}_{c}}{\dot{\rho}_{c}}\vec{v}+\dfrac{\vec{\nabla}\delta p}{\left(p_{c}+\rho_{c}\right)}+\vec{\nabla}\Psi.
%\label{eq:nu1}
%\end{equation}
This is the Euler equation for an ideal fluid in comoving coordinates. 

Some important remarks about \eqref{eq:nu1} are in order: we wish to analyze the matter perturbation of the universe, then, the term $\dfrac{\dot{p}_{c}}{\dot{\rho}_{c}}=c_s^2$, which corresponds to relativistic corrections to fluid velocity, is negligible. Furthermore, the gradient of the pressure fluctuations $\vec{\nabla}\delta p$, does not contribute to the matter content in the sub-horizon limit \cite{liddle2000,white20210}.
This model also consider the null pressure in the absence of perturbation, characterized by $w=0$.
Taking these considerations into account, the only relevant terms of equations \eqref{eq:nu0} and \eqref{eq:nu1} are
\begin{subequations}
\begin{align}
\dot{\delta}_{M} & =a^{-2}\vec{\nabla}\cdot\vec{v}-3\dot{\Phi}, \\
\dot{\vec{v}} & =\vec{\nabla}\Psi.
\end{align}
\end{subequations}
We can rewrite the above expressions in a more suitable form by differentiating the first equation with respect to the time and applying the divergence into the second one, so that
\begin{subequations}
\begin{align}
\ddot{\delta}_{M} & = a^{-2}\vec{\nabla}\cdot \dot{\vec{v}} - 2Ha^{-2}\vec{\nabla}\cdot \vec{v}-3\ddot{\Phi}, \label{eqdeltaM}\\
\vec{\nabla}\cdot\dot{\vec{v}} & =\nabla^{2}\Psi.
\end{align}
\end{subequations}
At last, at the sub-horizon scale $(k\gg aH)$ it is know that $\Phi\sim\cos\left(\dfrac{k}{aH}\right)$,
so, $\dot{\Phi}\approx\ddot{\Phi}\approx0$ \cite{liddle2000,white20210}.
Under this limit, we can eliminate $\vec{v}$ in \eqref{eqdeltaM} and find a differential equation for the contrast density of matter. 
Therefore, in the Fourier space we have
\begin{equation}
\ddot{\delta}_{M}+2H\dot{{\delta}}_{M}=-\dfrac{k^{2}}{a^{2}}\Psi.
\label{EDOdelta}
\end{equation}
In order to conclude the current analysis, it is convenient to write equation \eqref{EDOdelta} in terms of the $e$-folding time $N=\ln(a_0/a)$ and substitute $\Psi$ by the expression \eqref{eq:psi-1}, it results into
\begin{equation}
\delta_{M}^{\prime \prime } + \left( 2+\epsilon  \right)\delta_{M}^{\prime}-\dfrac{3H_{0}^{2}e^{3N}}{2H^2}\Omega_{M}^{0}\dfrac{\left(1+f+U_{c}+6V_{c}\right)}{\left(1+f+U_{c}\right)\left(1+f+U_{c}+2V_{c}\right)}\delta_{M}=0,
\label{EDOdeltaM}
\end{equation}	
in which  $\epsilon=H^{\prime} /H$ and the prime denotes differentiation with respect to $N$.
An interesting point is that this equation is  $k$ independent, depending only on the cosmological $e$-folding time $N$. 

\begin{figure}[H]
	\centering
	\includegraphics[scale=0.95]{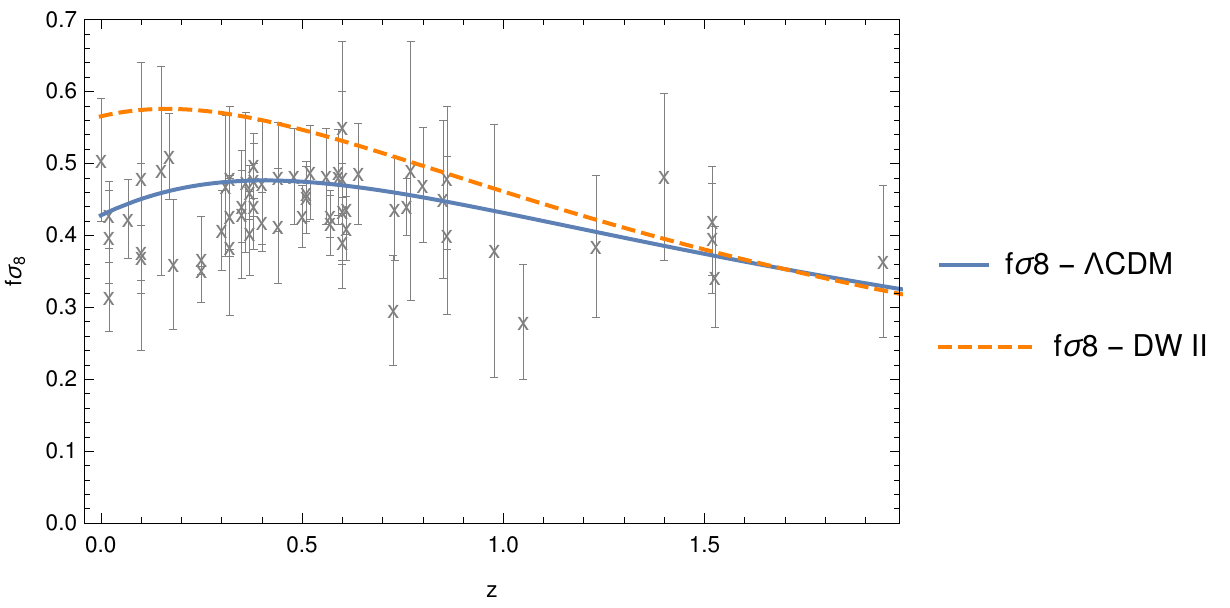}
	\caption{Comparison between the DW II and $\Lambda$CDM results for the $f\sigma_8$ observable in function of the cosmological red-shift $z$.}
	\label{fig:fsigma8}
\end{figure}

The product of the structural growth rate $f=\partial_N \ln [\delta_M]$ \footnote{Not to be confused with the distortion function $f(Y)$.}   and the amplitude of matter fluctuations in spheres of $8h^{-1}$Mpc, $\sigma_8=\sigma_8^0\frac{\delta_M(N)}{\delta_M(0)}$, is a physical observable related to the density contrast \cite{liddle2000,white20210}.
The value of the constant $\sigma_8^0=0.811$ has been recently constrained by the Plank satellite observations \cite{plank2018results}.
The numerical solution of $f\sigma_8$ in terms of the red-shift $z$, for the DW II and $\Lambda$CDM models, is presented in Figure \ref{fig:fsigma8} and compared with observational data \cite{ding2019structure}.
This analysis shows that we find that the $\Lambda$CDM solution is in good agreement with the data \cite{plank2018results}.
The analysis shows that the linear perturbation theory of the DW II model behaves regularly, and it is reliable and self-consistent as a whole.
Furthermore, unlike the result reported in \cite{ding2019structure}, there is no divergence in this observable for $0<z<2$. Although the $\Lambda$CDM model seems to be a best fit to the data rather than DW II, the non-local model cannot be ruled out by the observation of the growth rate $f\sigma_8$ at this redshift range.
It is important to remark that once the background is fixed to reproduce $\Lambda$CDM, no
more free parameters are left to adjust to the RSD data

%%%%%%%%%%%%%%%%%%%%%%%%%%%%%%%%%%%%%%%%%%%%%%%%%%%%%%%%%
%%%%%%%%%%%%%%%%%%%%%%%%%%%%%%%%%%%%%%%%%%%%%%%%%%%%%%%%%

\section{Structural Growth in Bouncing Cosmology}
\label{sec:grow}

We have finally reached our main analysis, which consists in examining early time perturbations for bouncing universes.
Our group have recently studied bouncing models in the context of non-local DW II cosmology \cite{jackson2022nonlocal}, where we have found that the reconstruction procedure generates physically consistent solutions to the distortion function for the following  bouncing solutions: symmetric bounce, oscillatory bounce, matter bounce, finite time singularity model and pre-inflationary asymmetric bounce.
Since the previous study was performed at the level of background cosmology, we seek to examine these models at perturbative level in order to further restrict physically relevant models.

\begin{figure}[tbp]
	\centering
	\begin{subfigure}[b]{0.32\textwidth}
		\centering
		\includegraphics[width=\textwidth]{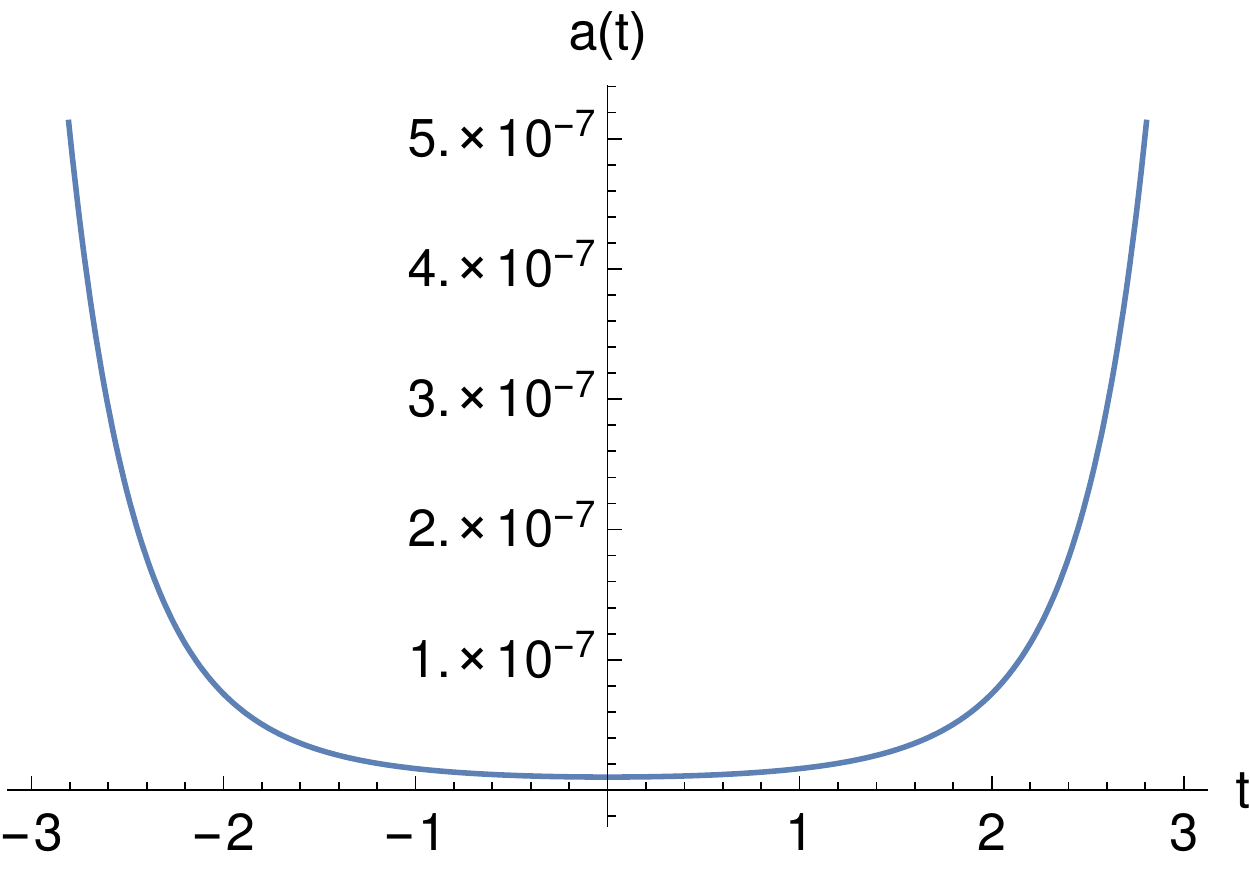}
		\caption{$a(t)=\frac{1}{10^{8}}e^{t^{2}/2}\,.$}
		\label{fig:aI}
	\end{subfigure}
	\hfill
	\begin{subfigure}[b]{0.32\textwidth}
		\centering
		\includegraphics[width=\textwidth]{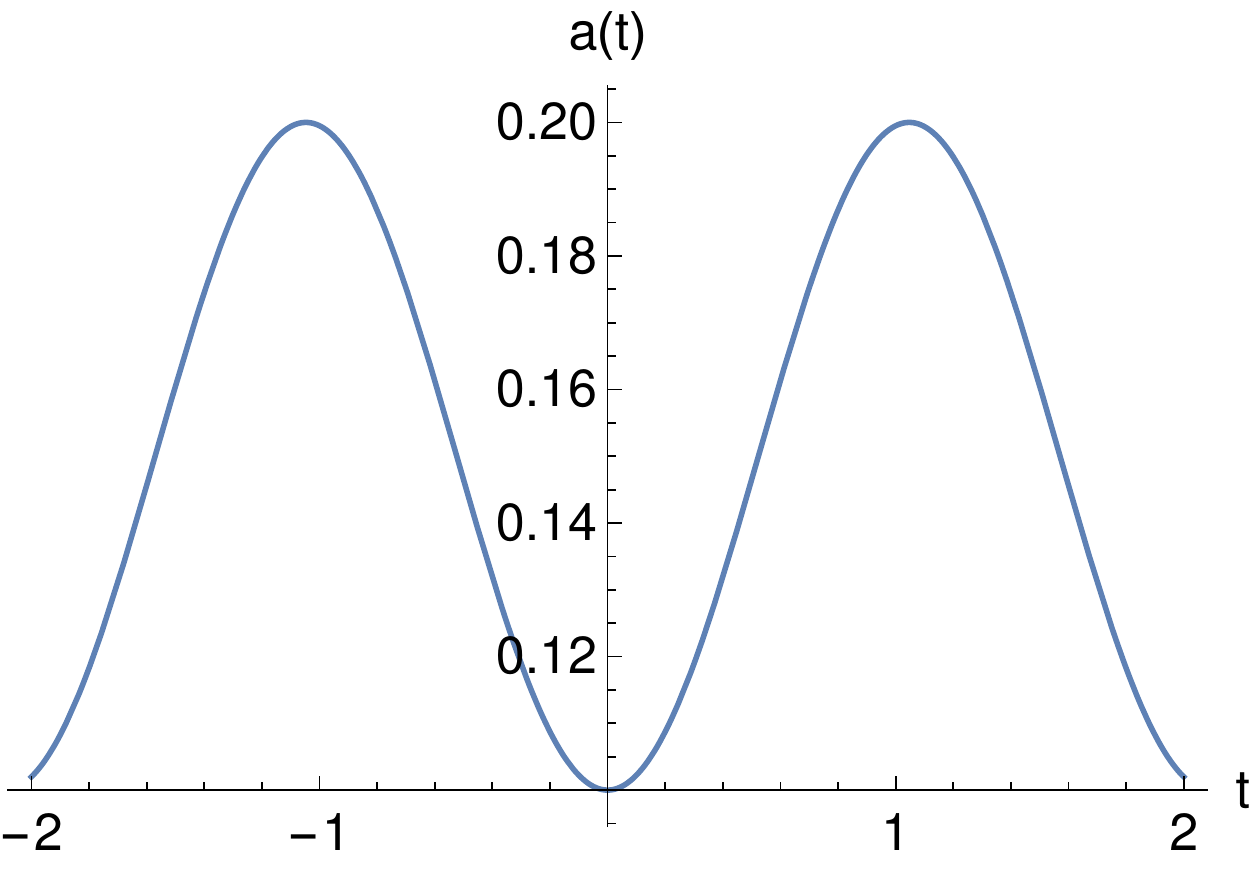}
		\caption{$a(t) = \frac{1}{10}[\sin^2{(\frac{3t}{2})}+1]$.}
		\label{fig:aII}
	\end{subfigure}
	\hfill
	\begin{subfigure}[b]{0.32\textwidth}
		\centering
		\includegraphics[width=\textwidth]{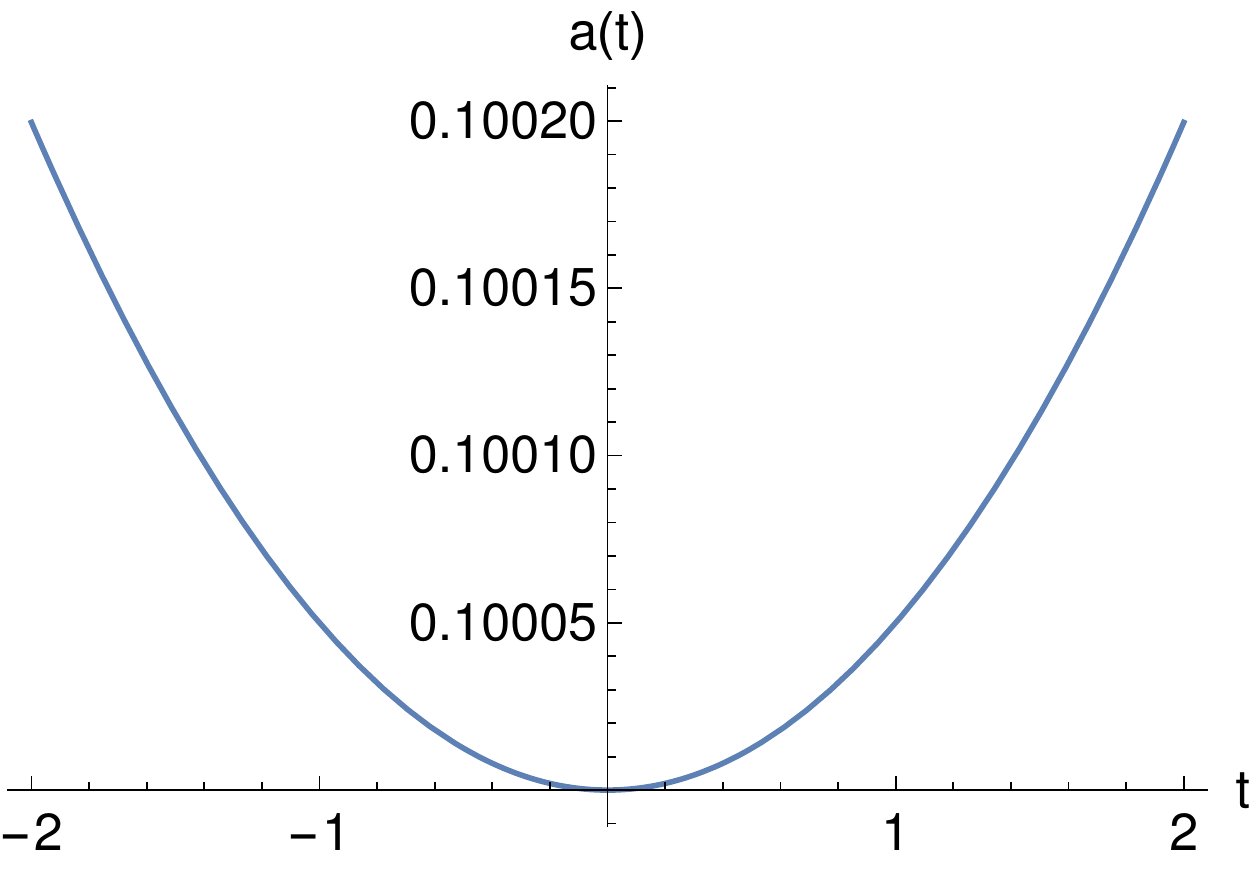}
		\caption{$a(t)=\frac{1}{10}\left( \frac{3}{2}\rho_c t^2+1 \right)^{1/3}$.}
		\label{fig:aIII}
	\end{subfigure}
	\hfill
	\begin{subfigure}[b]{0.32\textwidth}
		\centering
		\includegraphics[width=\textwidth]{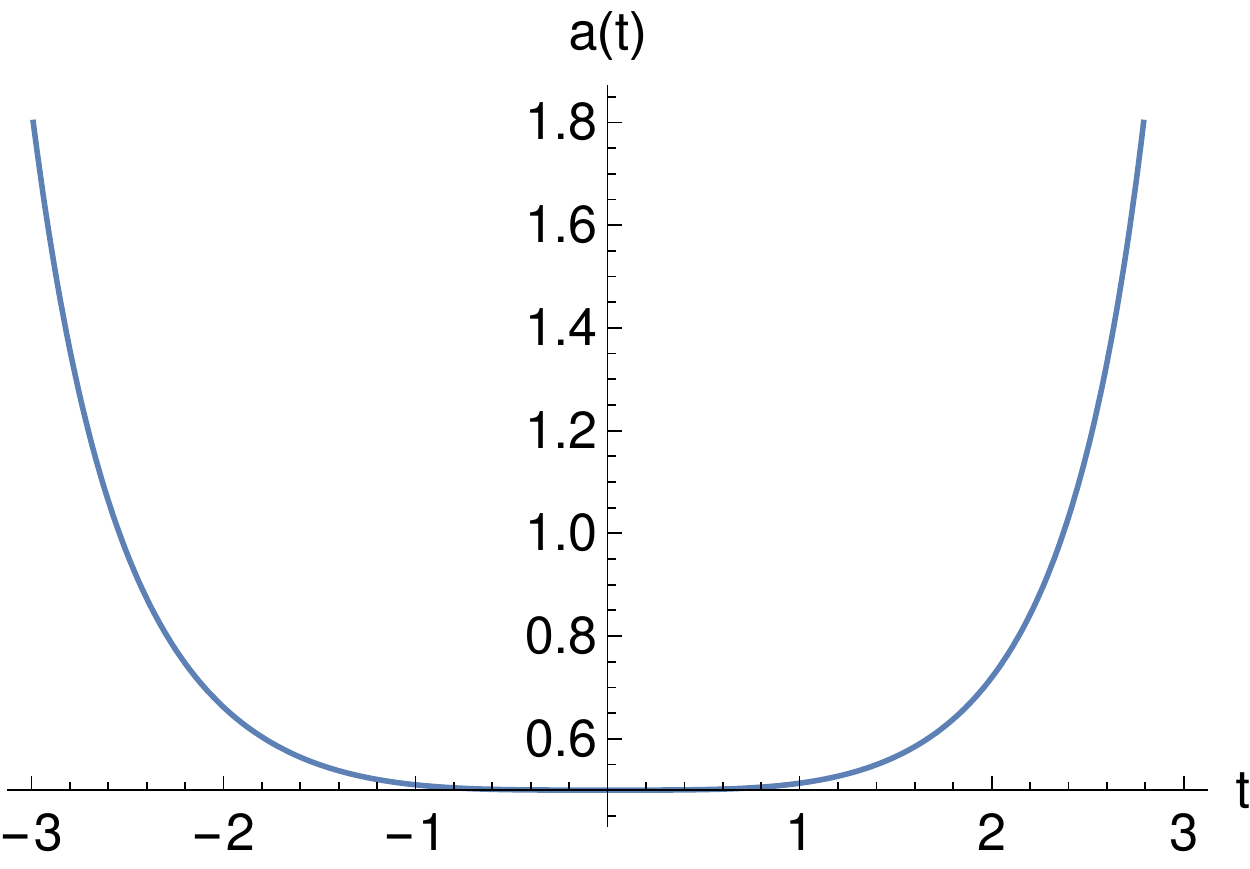}
		\caption{$a(t)=\frac{1}{2} e^{\frac{1}{10}\frac{t^{\alpha +1}}{\alpha +1}}$.}
		\label{fig:aIV}
	\end{subfigure}
	%\hfill
	\begin{subfigure}[b]{0.32\textwidth}
		\centering
		\includegraphics[width=\textwidth]{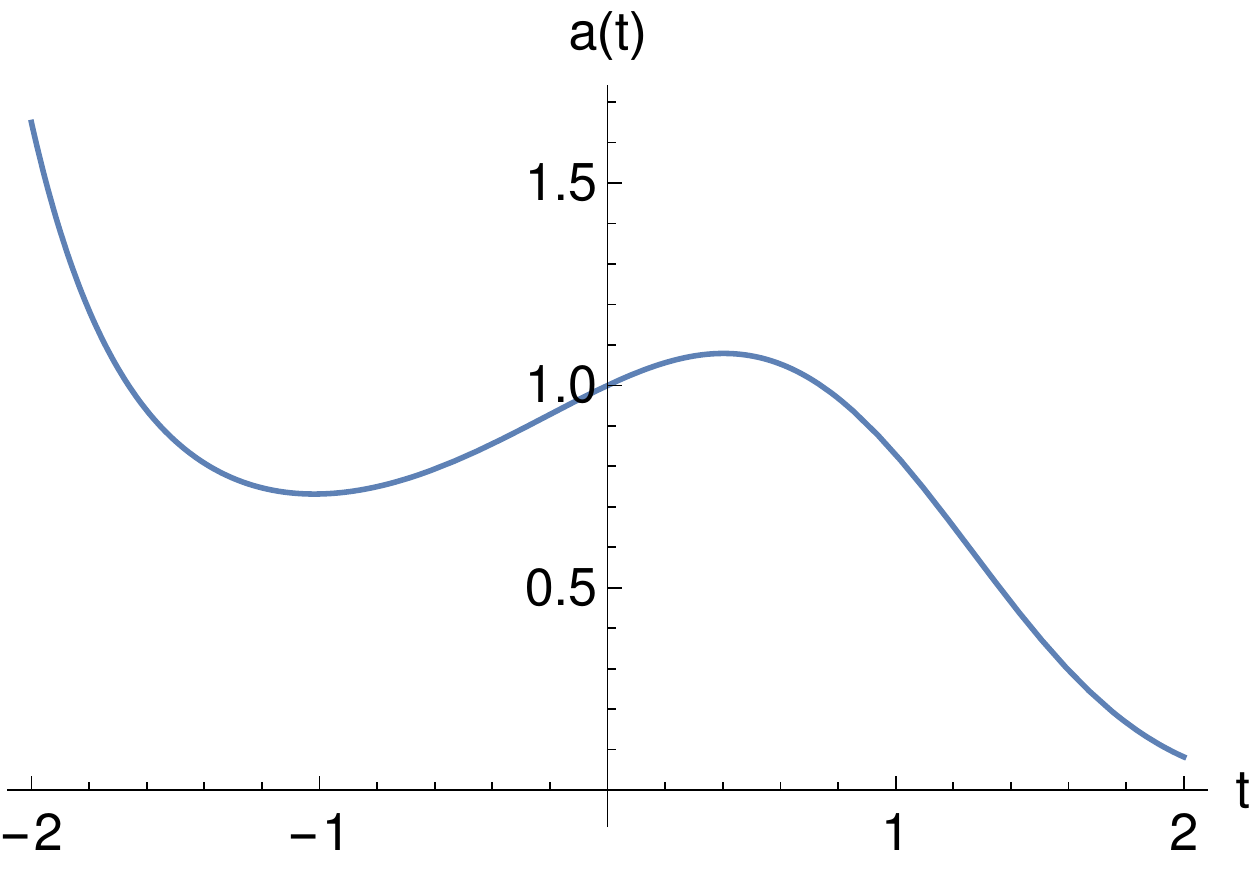}
		\caption{$a(t)=e^{-(\frac{11t}{17})^3-(\frac{t}{2})^2+\frac{t}{3}}$.}
		\label{fig:aV}
	\end{subfigure}
	\caption{The scale factor $a(t)$ for five different bouncing models: symmetric bounce, oscillatory bounce, matter bounce, finite time singularity model and pre-inflationary asymmetric bounce, respectively. In (c) the parameter $\rho_c \ll  1$ is the critical density and $\alpha >1$ in (d).}
	\label{fig:bouncing_a}
\end{figure}

We present next a brief review of each bouncing scenario, which are depicted in Fig.~\ref{fig:bouncing_a}:

\begin{enumerate}
\item The symmetric bounce model was initially proposed to study the $F(R)$ gravity \cite{Cai:2012va,Cai:2013vm,bamba2014bounce} since it generates a non-singular bounce and can be connected to the late time accelerated expansion.  It was also studied in the context of non-local DW I gravity \cite{chen2019primordial} and also in the $f(T,B)$ gravity \cite{caruana2020cosmological}.

\item The oscillatory bounce arises from the quasi-stead state cosmology \cite{novello2008bouncing,Cai:2012va,Steinhardt:2001st}, which was proposed as an alternative to the standard cosmology.
The oscillatory pattern of the scale factor was introduced to reproduce the cyclic behavior of the interchange between the domination of the cosmological constant and a scalar field with negative energy that create particles.

\item The matter bounce emerged in the context of loop quantum cosmology (LQC) \cite{Singh:2006im,Wilson-Ewing:2012lmx} and its scale factor satisfies the effective equations of LQC in the classical limit, for a dust-dominated universe.
These effective equations takes into account corrections due to quantum geometry into the usual Friedmann equations of the general relativity \cite{Singh:2006im,Wilson-Ewing:2012lmx,caruana2020cosmological}.

\item The bounce that generates finite time singularities is a more general exponential bouncing than the symmetrical bounce and was originally proposed to discuss the generation of singularities in the evolution of the universe \cite{Barrow:2015ora,Nojiri:2015fra,Odintsov:2015zza,Oikonomou:2015qha,caruana2020cosmological}. This model depends on the choice of the parameter $\alpha$, if $\alpha$ is chosen equal to $1$, it corresponds to the symmetric bounce. Moreover, the choice $\alpha = 0$ implies that the scale factor $a$ grows exponentially in time (de Sitter universe). Here we consider the case $\alpha > 1$ such that the scale factor and the effective energy density remains finite for every $t$.

\item In the pre-inflationary scenario, recently proposed in the $f(R)$ modified gravity \cite{odintsov2022preinflationary}, the universe contracts until it reaches a minimum size and  expands slowly entering a quasi de Sitter inflationary era. After that, the universe starts to contract again and the scale factor tends to zero. The motivation of this form of scale factor is that it avoids the cosmic singularity and approximately satisfies the String Theory scale factor duality condition $a(t) = a^{-1}(-t)$.
\end{enumerate}

The reconstruction procedure to encode the bouncing effects is analogous to the previously developed in section \ref{sec:imp}: first we should solve the equations for the auxiliary fields $X,Y,U,V,F$, then determine $f(t)$ from $f=F-1-U$.
Although, in this analysis we look for vacuum solution that reconstructs the bouncing evolution.
Thus, the differential equation for $F$ given by \eqref{eq:FNL} now becomes
\begin{equation}
\left[2\dot{H}+6H^{2}+d_t^{2}+5Hd_t\right]F(t)=0,
\label{eq:dif}
\end{equation}
which together with the solutions $f(t)$ and $Y(t)$ are used to obtain $f(Y)$. 

On the other hand, due to our purposes in studying bouncing universes, the density contrast $\delta_M$ differential equation \eqref{EDOdeltaM} is recast in terms of the time variable (instead of the e-folding time) as
\begin{align}
	\ddot{\delta}(t)+2H\dot{\delta}(t)-\frac{3H_{0}^{2}(F_c+8V_c)\Omega_{M}}{2a^{3}H^{2}(F_c(F_c+2V_c))}\delta(t)=0\,,
\label{eq:delta-b}
\end{align}
where $a(t)$ now assumes different forms for each bouncing cosmology (see Fig \ref{fig:bouncing_a}) and the subscript $c$ denotes that all the fields are only time dependent, since they are evaluated at the cosmological background.

\begin{figure}[tbp]
	\centering
	\begin{subfigure}[b]{0.45\textwidth}
		\centering
		\includegraphics[width=\textwidth]{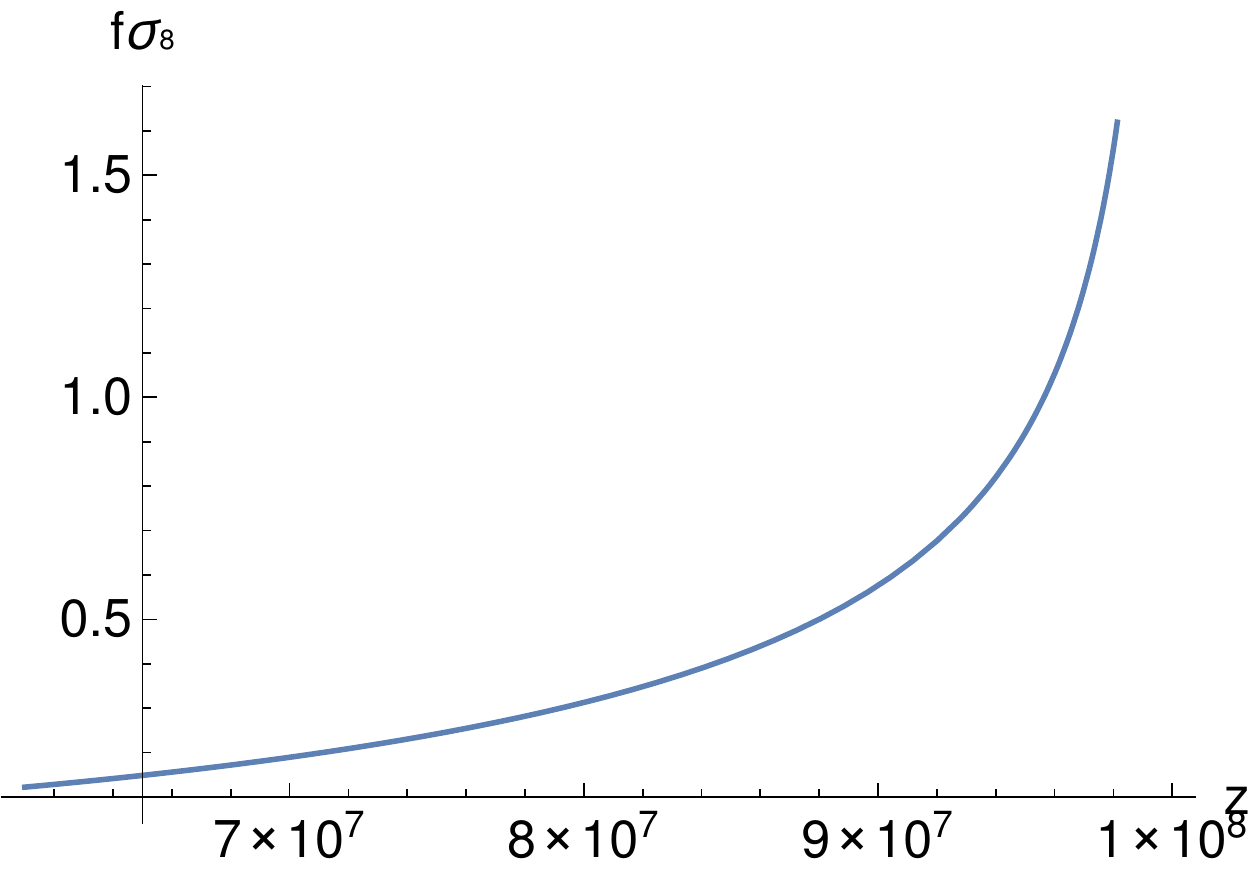}
		\caption{Symmetric bounce}
		\label{fig:fI}
	\end{subfigure}
	\hfill
	\begin{subfigure}[b]{0.45\textwidth}
		\centering
		\includegraphics[width=\textwidth]{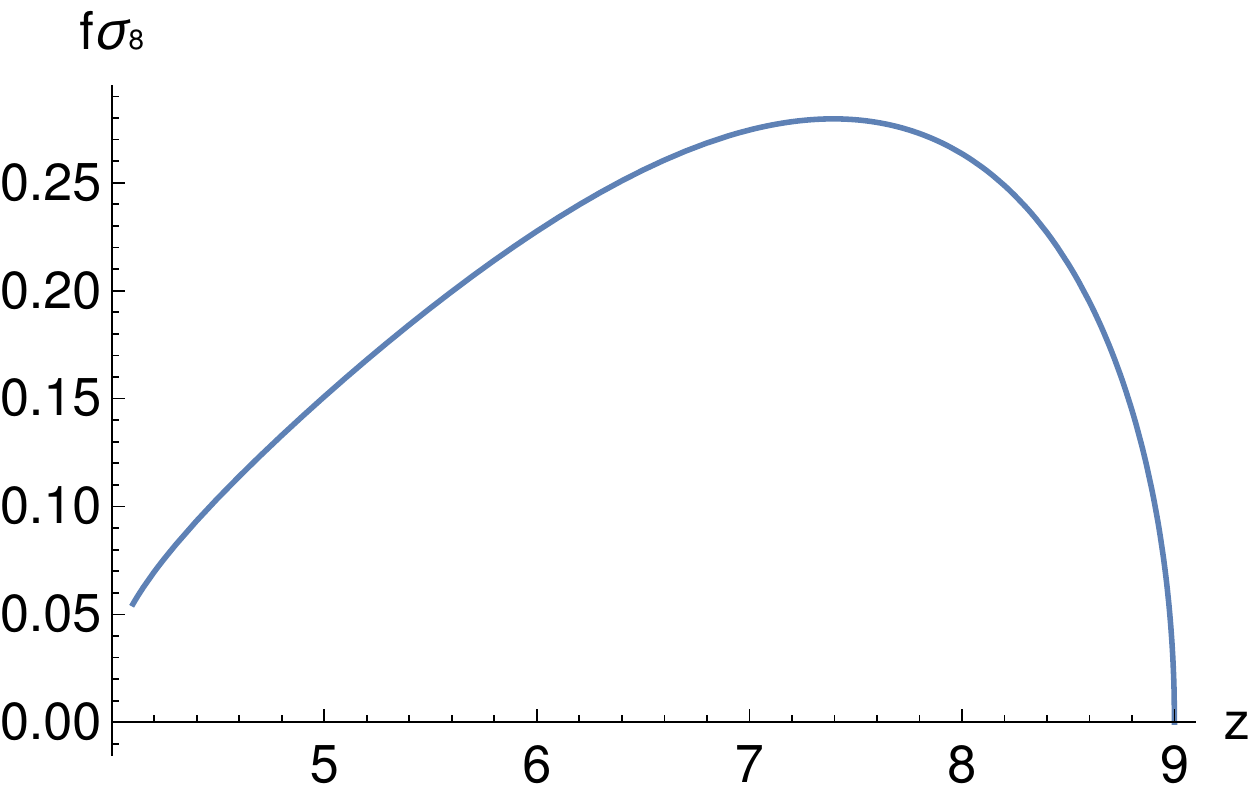}
		\caption{Oscillatory bounce}
		\label{fig:fII}
	\end{subfigure}
	\hfill
	\begin{subfigure}[b]{0.45\textwidth}
		\centering
		\includegraphics[width=\textwidth]{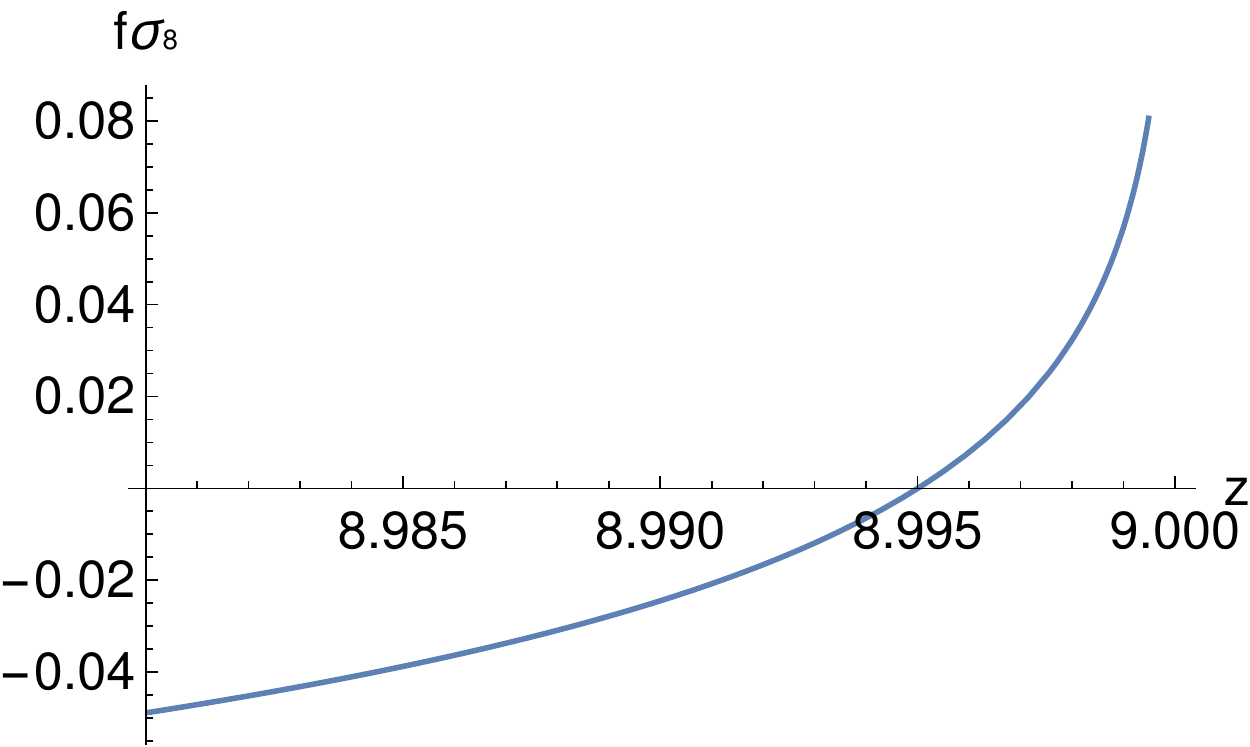}
		\caption{Matter bounce}
		\label{fig:fIII}
	\end{subfigure}
	%\hfill
	\begin{subfigure}[b]{0.45\textwidth}
		\centering
		\includegraphics[width=\textwidth]{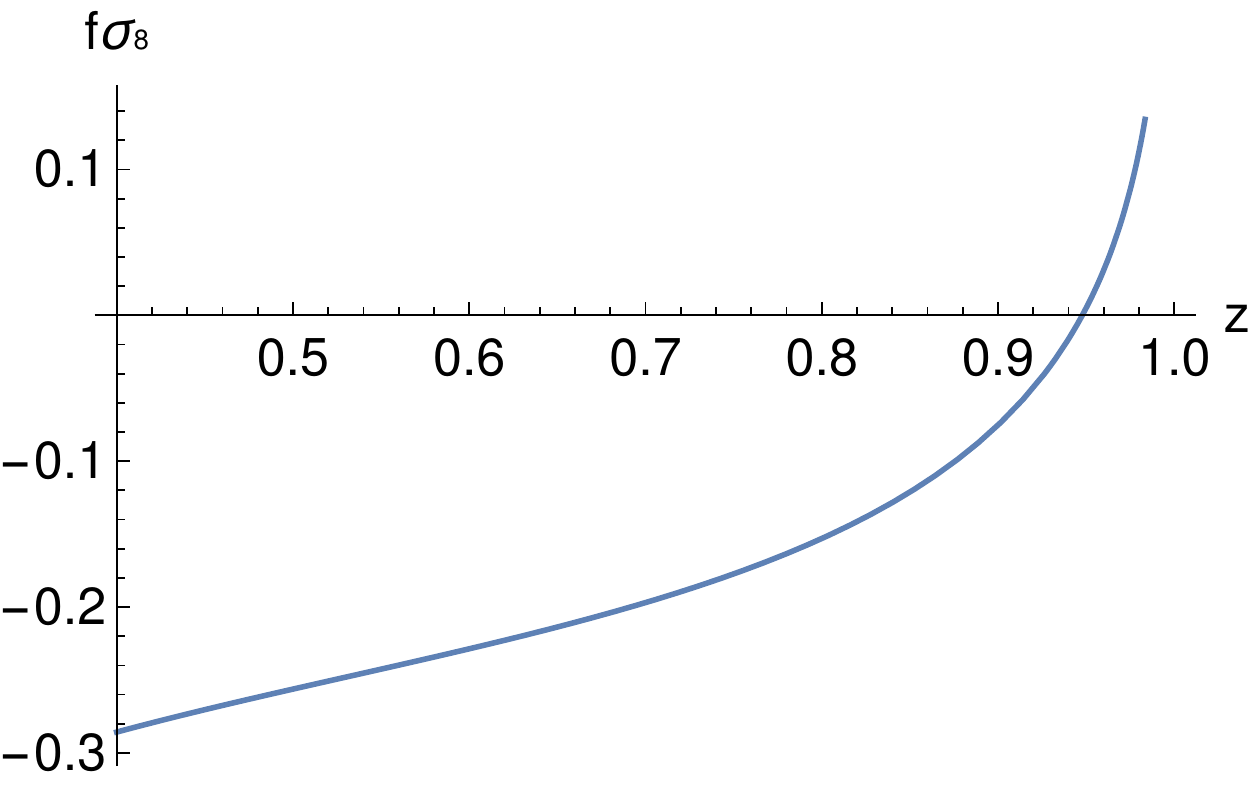}
		\caption{Finite time singularity}
		\label{fig:fIV}
	\end{subfigure}
	%\hfill
	\begin{subfigure}[b]{0.45\textwidth}
		\centering
		\includegraphics[width=\textwidth]{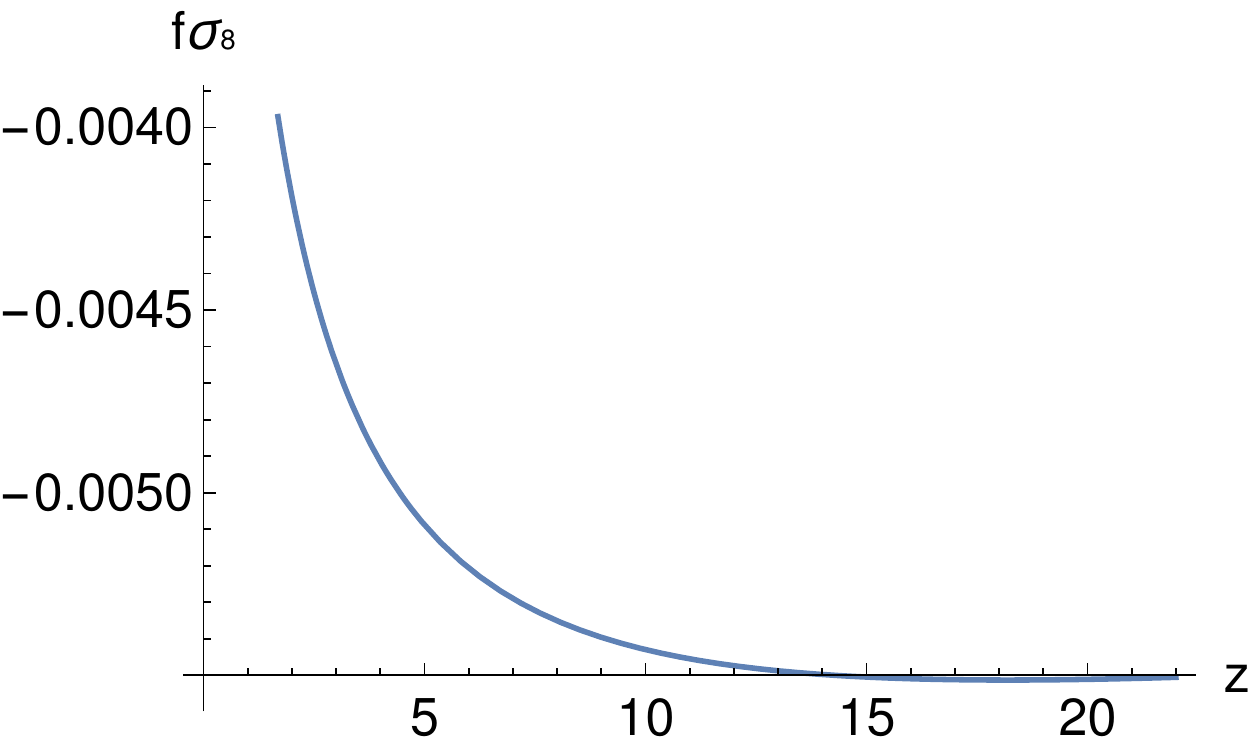}
		\caption{Asymmetric bounce}
		\label{fig:fV}
	\end{subfigure}
	\caption{The evolution of $f\sigma_8$ in terms of the redshift for five different bouncing models: symmetric bounce, oscillatory bounce, matter bounce, finite time singularity model and pre-inflationary asymmetric bounce, respectively.}
	\label{fig:bouncing_f}
\end{figure}

As before, expression \eqref{eq:delta-b} can be numerically solved for each bouncing scenario and used to evaluate the observable $f\sigma_8$.
We present the solution for each bouncing universe in terms of the cosmological redshift $z$ in Figure \ref{fig:bouncing_f}. Given the distinct behavior observed in the growth of structures across various bouncing cosmology scenarios, it is important to provide some remarks about our findings.

\begin{itemize}

\item Our calculation shows that in the symmetric bounce, matter bounce and finite time singularities, the observable $f\sigma_8$ has a growing pattern near the bouncing point (large $z$). This is a physically undesired effect since $f\sigma_8$ is a measure of the growth rate of matter 	perturbation in the early epoch and should approach a finite value as seen in the observational data depicted in Figure \ref{fig:fsigma8}.

\item  On the other hand, in the cases of oscillatory bounce and asymmetrical bounce, the matter fluctuations are very small at the bouncing point, as $f\sigma_8$ approaches zero for large redshift $z$, and the density contrast begins to grow immediately after the onset of expansion. This behavior may be regarded as the seeds, or fluctuations, that contribute to the formation of large scale structures in the universe. Thus, for an endlessly oscillating universe or a universe that starts with asymmetry, we can understand the origins of these fluctuations as non-local effects.

\end{itemize}

This analysis showed that the formation of the biggest structures currently observed in the universe cannot be described by the non-local Deser-Woodard II model in the case of the symmetric bounce, the bounce generated by critical matter density and the exponential bounce singular at a finite time.
In contrast, universes with oscillatory and pre-inflationary bounces may accomplish the formation of clusters of galaxies in the framework of DW II model.
Therefore, eternal universes with contractions and expansions in a nonlocal gravity model seems to be a better choice to describe the structure formation instead of models with a single minimum point.

\section{Conclusions}
\label{sec:conc}

In this paper we presented a comprehensive discussion of formation and growth of structures in the nonlocal Deser-Woodard II model in different bouncing cosmology scenarios. Initially, we revised the reconstruction process for the DW II model as well as the perturbation theory for the field equations. Next, we analyzed the perturbed nonlocal DW II model and its implications in some bouncing cosmologies, scrutinizing for physically acceptable solutions of the $f\sigma_8$ observable.

We began by revising the reconstruction process of the distortion function $f(Y)$ for the case of an accelerating expansion universe. During this analysis, we identified a small difference in the parameters of the exponential fit, compared to those reported by the original authors \cite{deser2019nonlocal}. This small difference is due to our choice of the right-hand side of equation \eqref{eq:difF} being derived directly from the equation \eqref{eq:rho} (compare it with equation (29) of \cite{deser2019nonlocal}). Although the parameters are different, the desired exponential growth is ensured in our solution for $f(Y)$, see equation \eqref{eq:fdeY}.

Furthermore, we analyzed the growth of structures in the universe by considering cosmological perturbations of the DW II model in the newtonian gauge. All field equations were expanded over a cosmological flat space FLRW background (time dependent only), with a small spatial dependence scalar perturbation. Naturally, the perfect fluid deviations of the stress energy tensor also were included. The field equations were evaluated in the sub-horizon limit, which provides a suitable way to study the matter density fluctuations. Our analysis shows that the structure growth rate $f\sigma_8$ is finite in the redshift range $0<z<2$, showing that the linear perturbation theory of the DW II model behaves regularly, and it is reliable and self-consistent as a whole. As we can see in Figure \ref{fig:fsigma8}, the experimental data favor the $\Lambda$CDM model while differing with the non-local DW II model curve.

As a complementary analysis, we have examined the formation of large scale structures by early time perturbations for different bouncing universes. This interest was motivated by the previous results where we have worked the physical viability of some bouncing cosmologies in the DW II model  \cite{jackson2022nonlocal}. Our analysis shows that the structure formation cannot be described by DW II model in the case of the symmetric bounce, matter bounce and the finite time singularity universe, as the observable $f\sigma_8$ presents an undesirable growing pattern near the bouncing point. On the other hand, the bouncing models with oscillations and the pre-inflationary bounce presented a physical behavior for the observable $f\sigma_8$. Thus, universes with successive contractions and expansions allows a description of the formation of large structures in term of non-local phenomena, instead of models with a single and finite bounce, at least in the particular framework that we have discussed.

 %%%%%%%%%%%%%%%%%%%%%%%%%%%%%%%
%\appendix
%\section{Some title}
%Please always give a title also for appendices.

\acknowledgments

This study was financed in part by the Coordenação de Aperfeiçoamento de Pessoal de Nível Superior - Brasil (CAPES) - Finance Code 001.
R.B. acknowledges partial support from Conselho
Nacional de Desenvolvimento Cient\'ifico e Tecnol\'ogico (CNPq Project No. 306769/2022-0).
%\paragraph{Note added.} This is also a good position for notes added
%after the paper has been written.


\end{document}